\documentclass[preprint,preprintnumbers,amsmath,amssymb,superscriptaddress,nofootinbib]{revtex4-1}

\usepackage{fullpage,amsthm,amsmath,amsfonts,bm,times,color,bbm}
\usepackage{mathtools}
\usepackage{amssymb}
\usepackage{graphicx}
\usepackage[english]{babel}
\usepackage{float}
\usepackage{caption}
\usepackage{subcaption}
\usepackage{multirow}
\usepackage{booktabs}
\usepackage{makecell}
\usepackage{wrapfig}
\usepackage{soul}
\usepackage{hyperref}
\usepackage{lineno}
\usepackage[title]{appendix}%
\usepackage{geometry}
\usepackage{booktabs}
\usepackage[table]{xcolor}

\begin{document}
\title{Planetary climate interactions of the Qinghai-Tibetan Plateau}
\author{Ziyan Wang}
\affiliation{School of Systems Science/Institute of Nonequilibrium Systems, Beijing Normal University, Beijing 100875, China}

\author{Teng Liu}
\affiliation{School of Systems Science/Institute of Nonequilibrium Systems, Beijing Normal University, Beijing 100875, China}
\affiliation{Earth
System Modelling, School of Engineering and Design, Technical University of Munich, Munich 80333, Germany}
\affiliation{Potsdam Institute for Climate Impact Research, Potsdam 14412, Germany}

\author{Shang Wang}
\affiliation{School of Systems Science/Institute of Nonequilibrium Systems, Beijing Normal University, Beijing 100875, China}

\author{Sheng Fang}
\affiliation{School of Systems Science/Institute of Nonequilibrium Systems, Beijing Normal University, Beijing 100875, China}

\author{Jun Meng}
\affiliation{State Key Laboratory of Earth System Numerical Modeling and Application, Institute of Atmospheric Physics, Chinese Academy of Sciences, Beijing, 100029, China}

\author{Xiaosong Chen}
\affiliation{School of Systems Science/Institute of Nonequilibrium Systems, Beijing Normal University,  Beijing 100875, China}

\author{J\"urgen Kurths}
\affiliation{Potsdam Institute for Climate Impact Research, Potsdam 14412, Germany}
\affiliation{Department of Physics, Humboldt University, Berlin 10099, Germany}

\author{Shlomo Havlin}
\affiliation{Department of Physics, Bar-Ilan University, Ramat Gan, 52900, Israel}

\author{Fahu Chen}
\affiliation{Institute of Tibetan Plateau Research, Chinese Academy of Sciences, Beijing 100101, China}

\author{Johan Rockstr\"om}
\affiliation{Potsdam Institute for Climate Impact Research, Potsdam 14412, Germany}

\author{Deliang Chen}
\affiliation{Department of Earth System Science, Tsinghua University, Beijing 100084, China}
\affiliation{Department of Earth Sciences, University of Gothenburg, Gothenburg 40530, Sweden}

\author{Hans Joachim Schellnhuber}
\affiliation{International Institute for Applied Systems Analysis, Laxenburg 2361, Austria}
\affiliation{Potsdam Institute for Climate Impact Research, Potsdam 14412, Germany}

\author{Jingfang Fan}
\email{jingfang@bnu.edu.cn}
\affiliation{School of Systems Science/Institute of Nonequilibrium Systems, Beijing Normal University, Beijing 100875, China}
\affiliation{Potsdam Institute for Climate Impact Research, Potsdam 14412, Germany}

\date{\today}
\clearpage
\begin{abstract}
The Qinghai–Tibetan Plateau (QTP), Earth's ``Third Pole", profoundly shapes the Asian monsoon and regional climate and exerts far-reaching influence on the global climate system. Yet its role in organizing planetary-scale climate interactions remains poorly quantified. Here we develop a climate network framework to explicitly resolve and quantify the planetary teleconnection architecture associated with the QTP based on historical observations and future climate projections, with physical consistency assessed using Lagrangian trajectory diagnostics and targeted numerical experiments. We uncover a persistent and directional interaction structure linking the QTP with multiple major climate tipping elements. In particular, we identify a robust tripolar interaction mode coupling the QTP with both the Arctic and Antarctica through coherent atmospheric–oceanic pathways. Our findings establish the QTP as a critical planetary climate integrator, revealing a significant blind spot in current climate models and risk frameworks regarding cascading tipping dynamics in a warming world.
\end{abstract}
\maketitle


Giant continental plateaus exert a profound influence on Earth’s climate by redistributing heat, momentum, and moisture within the atmosphere. Among them, the Qinghai–Tibetan Plateau (QTP) stands unrivaled in both scale and elevation, shaping large-scale circulation through intense surface heating, strong orographic forcing, and its central position within the Eurasian landmass~\cite{HuangGlobal2023}. Seminal studies have demonstrated that summer thermal forcing over the Plateau strengthens the South Asian High and drives the Asian monsoon, while winter mechanical forcing modulates midlatitude westerlies and cold-surge activity~\cite{BoosDominant2010}. Beyond these regional impacts, thermal anomalies over the QTP can excite stationary Rossby wave trains propagating along the westerly jet, enabling influences to extend far downstream~\cite{LiuPossible2017}. Diabatic heating over the Plateau further sustains large-scale vertical circulation cells that link the QTP with remote oceanic and continental regions~\cite{LuRole2018,NanVariability2019}. 

\begin{figure}[htbp]
  \vspace{-8mm}
  \centering
  \makebox[\textwidth][c]{\includegraphics[width=1.0\textwidth]{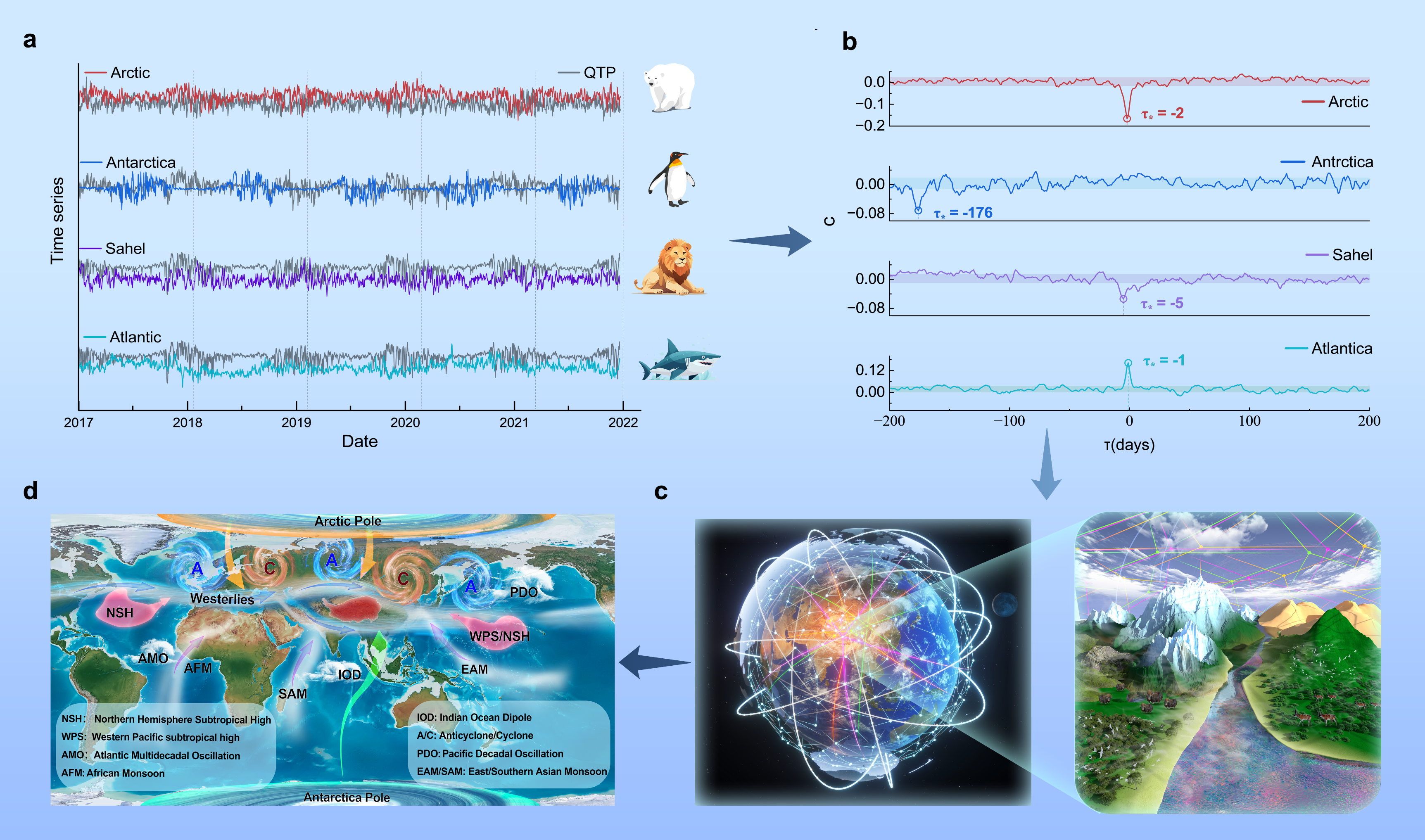}} 
  \captionsetup{labelformat=empty}
  \caption
  {\textbf{Fig.~1. Climate network identification of the Qinghai–Tibetan Plateau as a planetary-scale teleconnection nexus.} \textbf{a}, Daily near-surface temperature time series from representative nodes within the QTP and from selected remote regions, illustrating the variability used to infer long-range statistical interactions. \textbf{b}, Time-lagged cross-correlation function $C$ between QTP and external nodes [Eqs.~(\ref{EQ:C1}), (\ref{EQ:C2})]; the maximum absolute correlation defines interaction strength, while the sign of the corresponding lag $\tau_*$ determines the direction of influence. Shading indicates the range of the mean ± 3 standard deviations of $C$. \textbf{c}, Schematic representation of the climate network, in which colored links denote statistically significant and directional teleconnections connecting the QTP to other regions. The inset highlights the QTP as a multicomponent Earth-system entity integrating atmospheric, cryospheric, hydrological, and biospheric processes. \textbf{d}, Conceptual summary of the physical processes through which the QTP interacts with the global climate system, including diabatic heating, orographic forcing and wave propagation, providing the physical basis for the network-based detection of planetary-scale interactions.  
\label{Fig1:intro}
    }
\end{figure}

Despite decades of progress, investigations of QTP-related climate interactions have remained largely fragmented. Existing studies have predominantly focused on (i) localized or regional-scale couplings, (ii) the long-term evolution of Plateau–climate interactions inferred from paleoclimate records~\cite{YangRole2024,YangPortraying2020}, or (iii) the climatic impacts of individual surface processes, such as Plateau surface darkening or land–atmosphere feedbacks~\cite{TangResonance2024}. As a result, the system-level architecture through which the QTP organizes climate interactions across spatial scales, timescales, and physical mechanisms has yet to be understood. Addressing this gap is increasingly urgent. The QTP constitutes a tightly coupled Earth system in which atmospheric circulation, snow and ice processes, terrestrial ecosystems, and hydrological dynamics interact nonlinearly (Fig.~\ref{Fig1:intro}c). At the same time, it has emerged as one of the most climate-sensitive regions globally, experiencing amplified warming, accelerated glacier retreat, and widespread greening over recent decades~\cite{HussGlobalscale2018,KaabMassive2018,PiaoImpacts2012,ZhuGreening2016}. These rapid transformations raise the possibility that the QTP may act as a climate tipping element, capable of amplifying disturbances and transmitting them across the climate system through cascading feedbacks~\cite{LiuTeleconnections2023,YaoRecent2019,SunRegional2023}. Because tipping elements can generate impacts far beyond their region of origin, triggering abrupt and potentially irreversible changes elsewhere~\cite{WunderlingClimate2024,ArmstrongMcKayExceeding2022}, determining whether, and by what mechanisms, the QTP interacts dynamically with other rapidly changing regions is essential for understanding global climate stability and associated risks under ongoing warming.

Here we apply a climate network (CN) framework grounded in statistical physics and complexity theory~\cite{NewmanNetworks2010,BoersComplex2019,DijkstraNetworks2019,LudescherNetworkbased2021,FanStatistical2021,MengArctic2023} to map the QTP's planetary teleconnection structure using both reanalysis data and CMIP6 simulations. The network reveals a spectrum of persistent and bidirectional linkages connecting the QTP with multiple climate tipping elements, including the Arctic, Antarctica, ENSO, the Sahel, the tropical Atlantic, and the Amazon. Among these, a striking \textbf{tripolar mode} emerges that couples the QTP with both poles through coherent atmospheric and oceanic pathways. Validated by Lagrangian trajectory analyses and climate model experiments, this structure establishes the QTP as a planetary climate nexus and a missing integrator in global climate modeling and risk assessment (Fig.~\ref{Fig1:intro}d).

\section*{Results}
\subsection*{Planetary interaction pattern of the QTP}

\noindent To unveil the interactions between the QTP and the global climate system, we divide the CN nodes into two communities: those located within and outside the QTP region, and construct directed networks between the two for each year. Edges are defined as the maximum absolute time-lagged Pearson correlations between daily near-surface temperature time series of node pairs across the two communities. The corresponding time lag determines the direction of each edge, and only significant edges were retained (Fig.~S1d). Representative examples of such connections between the QTP and the Arctic, Antarctica, Sahel, and tropical Atlantic are shown in Fig.~\ref{Fig1:intro}a and b, along with their associated time series and correlation functions. For each external node $j$ and year $y$, we define its out-degree $D_{\mathrm{out}}^y(j)$ and in-degree $D_{\mathrm{in}}^y(j)$ (Eqs.~(\ref{EQ:outdegree}) and~(\ref{EQ:indegree}); Fig.~S1e). To identify robust and stable interaction patterns, we calculate the frequency with which each node ranks within the top 5\% of degree values across years, denoted as $F_{\mathrm{out}}$ and $F_{\mathrm{in}}$ (collectively referred to as $F$; Eqs.~(\ref{EQ:F_OUT}) and~(\ref{EQ:F_IN})), which quantify the \textbf{directional strength of influence} toward and from the QTP, respectively.

Statistical significance testing against a shuffle-based null model confirms that these connections are not random artifacts (Methods and Extended Data Fig. 1a). Notably, the distributions of $F_{\mathrm{out}}$ and $F_{\mathrm{in}}$ display pronounced peaks at large distances, indicating that the QTP is not confined to regional influence but participates in organizing climate interactions across hemispheres. The resulting \textbf{planetary-scale interaction patterns} are shown in Fig.~\ref{Fig2:pattern}, which reveal a highly structured teleconnection architecture that exhibits persistent, high-frequency coupling with the QTP in several key regions.

For adjacent regions, elevated $F_{\mathrm{out}}$ values west of the QTP (Fig.~\ref{Fig2:pattern}a, rectangle \uppercase\expandafter{\romannumeral2}), encompassing Central Asia, northwestern India, and the northwest Arabian Sea, align with the well-recognized influence of midlatitude westerlies, reinforcing the Plateau’s role as a hub for extratropical circulation interactions. High $F_{\mathrm{in}}$ values over East and South Asia (Fig.~\ref{Fig2:pattern}b, rectangle \uppercase\expandafter{\romannumeral5}) further corroborate the QTP’s established regulatory role in the Asian monsoon system~\cite{WuTibetan1998}.

A striking feature is the spatial congruence between the QTP and recognized global climate tipping elements~\cite{LentonClimate2019,ArmstrongMcKayExceeding2022}. The QTP maintains a robust dialogue with the Amazon and the Sahel~\cite{LiuTeleconnections2023,TrauthEarly2024,ChenImpact2021,NanVariability2019}. Besides, there are also strong and persistent bidirectional couplings with ENSO and the tropical Atlantic, the latter being associate with the famous tipping element, the Atlantic Meridional Overturning Circulation (AMOC)~\cite{FanNetwork2017,ArmstrongMcKayExceeding2022,CaesarObserved2018,LohmannMultistability2024}. This interaction reflects the QTP’s active role within the coupled atmosphere–ocean system, whereby ENSO and Atlantic not only influence the QTP, but Plateau thermal and mechanical forcing can modulate large-scale circulation patterns that influence tropical Pacific and Atlantic variability in modern era~\cite{XieOceanic2023,YuPotential2023,SunInterannual2019}.

The most prominent feature is the explicit identification of \textbf{tripolar teleconnections} that couple the QTP with both the Arctic and Antarctica (Fig.~\ref{Fig2:pattern}c-e), providing direct empirical support for a mechanism that has long been hypothesized but lacked quantitative confirmation~\cite{DuanClimate2025}. In the Arctic (Fig.~\ref{Fig2:pattern}c), inner seas including the Barents–Kara Seas and the Beaufort–Chukchi–East Siberian–Laptev sector exert a pronounced influence on the QTP. This influence may be modulated by the Arctic Oscillation~\cite{MengArctic2023,ZhangArcticTibetan2019}, as well as by reduced surface albedo and enhanced lower-tropospheric heat fluxes associated with sea-ice loss~\cite{DuanSea2022,LiArctic2020}. Conversely, the QTP exerts a discernible influence on Arctic regions, including the Barents–Kara Sea, Baffin Bay, the Bering Sea, and the Sea of Okhotsk (Fig.~\ref{Fig2:pattern}e), through snow cover anomalies~\cite{ZhangDynamic2023}. Along the Antarctic margin, the Amundsen Sea, Bellingshausen Sea, and Haakon VII Sea, including the Lazarev, Riiser Larsen, and Cosmonaut Seas, also exert a detectable influence on the QTP (Fig.~\ref{Fig2:pattern}d). This influence is likely mediated by the Antarctic Oscillation~\cite{TangSurface2022} and alters thermal conditions over the QTP through large-scale air–sea interactions~\cite{LiangAsian2022}.

Collectively, these results define a previously unrecognized planetary-scale interaction structure, the \textbf{QTP-Mode}. The structural stability of the QTP-Mode was validated through sensitivity tests involving a spectrum of alternative thresholds, ranging from initial edge detection to degree distributions and frequency filter, as well as variable substitution using 1000 hPa temperature data (Figs.~S2–S4).

To rigorously verify the physical connectivity of the QTP-Mode, we performed causality analyses between the QTP and the identified remote key regions. First, representative time series were constructed by averaging the grid points exhibiting significant $F$ intensity, and their stationarity was strictly confirmed. Two complementary approaches were employed: Granger causality analysis~\cite{Granger1969Investigating} and the Liang-Kleeman Information Flow (LKIF) method~\cite{Liang2005Information, Liang2014Unraveling}. The results reveal complex causal architectures (Extended Data Table~1). Specifically, the interaction with Antarctica exhibits a distinct asymmetry, as the LKIF analysis detected a prominent unidirectional information flow from Antarctica to the QTP with no significant feedback loop, a finding that strongly corroborates our conclusion. Ultimately, these causal pathways substantiate the QTP-Mode as a physically coupled structure.

While individual pairwise linkages between the QTP and some of these regions have been documented previously, our results for the first time uncover their systematic organization into a coherent planetary interaction framework. Patterns demonstrate that these teleconnections are largely bidirectional and persistent over decades, forming a stable backbone of the QTP teleconnection. This bridges fragmented regional feedbacks and fills a critical gap in understanding system-level climate teleconnections.

\subsection*{Propagation pathways and physical mechanisms of QTP teleconnections}

\noindent The QTP-Mode (Fig.~\ref{Fig3:paths}a), characterized by persistent and directional teleconnections linking the QTP with major climate tipping elements, including the Amazon, Sahel, AMOC, ENSO and the polar regions (Arctic and Antarctica), represents a dynamically coherent planetary interaction structure. However, the persistent propagation pathways and underlying dynamical mechanisms that sustain this mode remain unclear. These linkages enable cross-hemispheric transfer of energy and information through atmospheric and oceanic waveguides, shaping the spatial organization and temporal evolution of climate anomalies~\cite{LiEquatorial2019,LiuAtmospheric2007}. Resolving their propagation pathways is therefore crucial for understanding not only present climate variability but also the potential for cascading climate responses~\cite{WunderlingClimate2024,CohenRecent2014}.

To map these pathways, we apply an unpartitioned CN using 6,570 quasi-uniformly distributed grid points as nodes (see Methods). Using a minimal-cost function combined with shortest-path algorithms~\cite{ZhouTeleconnection2015,LiuTeleconnections2023}, we identify physically plausible propagation routes embedded within the network topology, revealing both canonical and previously unresolved teleconnections between QTP and key climate zones, particularly its tripolar connectivity.

In the Northern Hemisphere (Fig.~\ref{Fig3:paths}b), Arctic sea-ice loss or a positive phase of the Arctic Oscillation can initiate meridional Rossby wave trains that propagate southward through the Siberian High and modulate thermal and dynamical anomalies over the QTP~\cite{DuanSea2022,MengArctic2023}. Conversely, feedback from the QTP toward the Arctic is primarily transmitted zonally by the midlatitude westerlies into the North Pacific~\cite{XiaoImpacts2016}, where interactions with the Aleutian Low redirect the flow poleward, re-coupling the signal to the Arctic circulation~\cite{OverlandDecadal1999}.

In the Southern Hemisphere (Fig.~\ref{Fig3:paths}c), the coupling between West Antarctica and the QTP is likely mediated by variability in the Antarctic Oscillation and evolving sea-ice conditions~\cite{RogersSpatial1982, GongDefinition1999}. Wave energy generated over the Amundsen and Haakon VII Seas propagates eastward within the Southern Hemisphere westerlies~\cite{NicholsonITCZ2018}, interacts with the Intertropical Convergence Zone and African monsoon systems, and is subsequently steered northeastward toward the QTP by the Northern Hemisphere westerlies~\cite{KongInteraction2020}.

To reveal the physical realism of these network-derived pathways, we conduct ensemble simulations using the Hybrid Single-Particle Lagrangian Integrated Trajectory (HYSPLIT) model~\cite{SteinNOAAs2015}(see Methods),
and examine the steering influence of large-scale atmospheric circulation.
Strikingly, the simulated particle trajectories (yellow, Fig.~\ref{Fig3:paths}b–c) closely align with the network-inferred paths (red), supporting the existence of preferred atmospheric corridors for long-range coupling. Figure S5 presents representative ensemble structures. This geometric realism is further substantiated by the 500 hPa wind fields (Fig.~S7), which show that the identified teleconnection corridors, particularly the tripolar loop, are highly consistent with the background wind climatology and anomalous steering flows across different decades. The convergence across network topology, model transport, and wind diagnostics enhances confidence in both the geometric realism and dynamical plausibility of the  propagation patterns.

Beyond polar coupling, additional teleconnection pathways connect the QTP with tropical and subtropical key regions (Extended Data Fig.~2). The Sahel-QTP pathway (Extended Data Fig.~2a) follows a sequence involving vertical motion over the Plateau, descent toward South Asia, and advection into West Africa via the northeast trade winds. QTP–ENSO and bidirectional QTP–Atlantic pathways (Extended Data Fig.~2b-d) are primarily routed along the westerly jet, whereas the Atlantic-QTP route takes a westward detour around the Gulf stream before proceeding westward along the jet. In contrast, ENSO–QTP coupling (Fig.~S8e–l) exhibits multiple waveguides involving both eastward and westward propagation across the Pacific basin and Eurasia continent~\cite{CaiChanging2021,CaiIncreased2018,GengIncreased2023, WangHistorical2019}. The consistency between the network-derived pathways and the 500 hPa wind vectors (Fig.~S9) provides additional empirical basis for the identified teleconnections.

All pathway robustness is also assessed through repeated analyses using alternative source and target locations (Figs.~S10–S14) and extended to CMIP6 historical simulations (1950–2015; Fig.~\ref{Fig3:paths}d,e and Fig.~S15), supporting the temporal stability and spatial coherence of the identified routes. Interestingly, propagation times inferred from both network and HYSPLIT analyses (Fig.~\ref{Fig3:paths}f; Table~S2,3) reveal pronounced asymmetry, with characteristic transit scales of approximately 7 days for Arctic-to-QTP transmission, 23 days for QTP-to-Arctic feedback, and 25 days for Antarctica-to-QTP pathways, consistent with synoptic-scale atmospheric dynamics~\cite{LiuAtmospheric2007}. Repeating the analysis with 1000 hPa temperature yields comparable results (Figs.~S16 and 17), supporting variable independence.

To disentangle the relative roles of dynamical and thermal forcing in shaping the QTP-Mode, we conduct targeted numerical experiments that separately modified the Plateau’s mechanical and thermal influences by removing topography (amip\underline{~~}NMO) and suppressing near-surface diabatic heating (amip\underline{~~}NS\underline{~~}MO) over the QTP and adjacent regions (see Methods). While the control simulation reproduces the characteristic QTP-Mode (Fig.~S19a,b), both perturbations substantially enhance local interactions near the Plateau while significantly weakening large-scale teleconnections to remote regions (Fig.~S19c-f; Table~S3). The stronger suppression under amip\underline{~~}NMO highlights the dominant role of mechanical forcing in establishing planetary-scale waveguides and organizing stationary Rossby wave trains. In contrast, the amip\underline{~~}NS\underline{~~}MO response demonstrates that diabatic heating provides an essential thermodynamic contribution by amplifying and sustaining long-range coupling. These experiments indicate that the QTP-Mode emerges from the coupled action of topographic dynamics and thermal forcing, neither of which alone is sufficient to maintain its planetary-scale structure.

\subsection*{Tripolar teleconnections under climate change scenarios}

\noindent The CN and HYSPLIT analyses consistently reveal robust and physically interpretable tripolar teleconnections linking the QTP with the Arctic and Antarctica. In particular, the directional influence of the QTP to Antarctica and the bidirectional coupling between the QTP and the Arctic emerge as structurally stable components of the QTP-Mode. However, these teleconnections are embedded in regions undergoing rapid climatic change. Under ongoing warming, the polar components of the Earth system have experienced widespread retreat, with anthropogenic warming expected to be especially pronounced in high-latitude regions~\cite{DingGlobal2019}.
The QTP is warming at approximately twice the global average~\cite{YaoRecent2019}, and its snow cover has shown signs of approaching a tipping point since around 2008~\cite{LiuTeleconnections2023}. At the same time, the Arctic sea ice and the Greenland Ice Sheet are undergoing accelerated decline~\cite{ScreenCentral2010}, while Antarctica and the Southern Ocean exhibit sustained changes in ice mass and ocean heat content~\cite{JoseyRecordlow2024,CasadoQuandary2023}. These interlinked changes across the three polar regions have the potential not only to reshape regional thermodynamic gradients but also to reorganize long-range teleconnections, with far-reaching consequences for hydrology, ecosystem stability, and the planetary energy balance~\cite{EnglandTropical2020,ZhaoGlobal2019}.

To evaluate the evolution of the tripolar teleconnections under projected climate forcing, we analyze an ensemble of CMIP6 models (Table~S1) following the high-emission Shared Socioeconomic Pathway (SSP 5-8.5) scenario for 2015–2100. This period is divided into two 43-year epochs (2015–2057 and 2058–2100). For each model and epoch, we replicate the CN framework to derive model-specific $F_{\mathrm{out}}$ and $F_{\mathrm{in}}$ fields (see Methods). Representative examples and ensemble-mean patterns are shown in Fig.~S18-20. Aggregating across all models yields multi-model ensemble frequencies, $FM_{\mathrm{out}}$ and $FM_{\mathrm{in}}$ [Eqs.~(\ref{EQ:FModel_OUT}) and (\ref{EQ:FModel_IN})], which quantify the inter-model consistency of directional teleconnections across the ensemble (Fig.~\ref{Fig4:FuturePattern}a–d).

Across both the near-future (2015–2057) and late-century (2058–2100) periods, results show that the Southern Ocean is projected to maintain or even strengthen its influence on the QTP (Fig.~\ref{Fig4:FuturePattern}a,b), while Arctic seas, particularly the Beaufort, Chukchi, and East Siberian sectors, increasingly exhibit sensitivity to QTP-driven influence (Fig.~\ref{Fig4:FuturePattern}c,d).

Differences in ensemble frequencies between the two epochs (Extended Data Fig.~3) further reveal a spatial reorganization of these teleconnections. Southern Ocean influence intensifies over the outer Weddell Sea and the interior Amundsen and Ross Seas, while weakening along the zonal band near 60°S. Simultaneously, the QTP’s impact strengthens over the central Arctic, spanning the Beaufort to East Siberian seas, with weaker responses in marginal regions such as the Barents Sea. These changes indicate a poleward shift and increased spatial focusing of cryosphere-linked teleconnections, likely shaped by evolving sea-ice conditions and storm-track dynamics.
 
Crucially, this spatial realignment closely correlates with projected sea-ice retreat. Across models and decades, winter sea-ice boundaries consistently coincide with regions where changes in teleconnection strength are most pronounced (Extended Data Fig.~3). This correspondence indicates that cryospheric change is not merely a passive background response to warming, but an active dynamical agent shaping the architecture of long-range climate interactions. As warming progresses, zones of sea-ice retreat may increasingly function as both amplifiers and gateways for atmospheric signal transmission, thereby reconfiguring the topology of global climate connectivity. The persistence of these tripolar linkages is further supported by parallel analyses using 1000 hPa temperature as an alternative variable (Fig.~S22) and by robust tripolar propagation pathways shown in Extended Data Fig.~4 and Fig.~S23, confirming the structural stability of the QTP-Mode under variable substitution.

The "thermo-dynamical amplifier" mechanism over the QTP may re-energize polar signals which prone to dissipation across the equatorial zone. This process facilitates a "Cryospheric Resonance," may characterized by synchronized surface darkening across the Three Poles. The resulting synergistic reduction in global albedo exacerbates the planetary energy imbalance, accelerating simultaneous melting and establishing a cross-hemispheric "melt-warm-melt" feedback loop. As a coupled system, the critical threshold is likely determined by the "weakest link," marking the onset of a transition from linear response to nonlinear amplification. This synchronized destabilization may pose severe risks, including the accelerated collapse of the global cryosphere and profound shifts in the Earth's energy budget. Regionally, the alteration of QTP thermal forcing threatens to disrupt the Asian monsoon system, while teleconnected atmospheric bridging may increase the concurrency of extreme weather events across hemispheres. Due to the significant inertia inherent in ocean-cryosphere interactions, the reversibility of this system is assessed as low.

\section*{Discussion}

\noindent The QTP has traditionally been viewed as a regional regulator of the Asian monsoon system~\cite{WuTibetan1998}. Our results show that this perspective substantially underestimates its role in the Earth system. By integrating CN analysis with Lagrangian trajectory diagnostics, we reveal here that the QTP functions as a planetary-scale climate integrator, persistently and directionally coupled to both polar regions and to multiple major climate tipping elements. This coupled structure defines a dynamically stable and physically coherent global interaction pattern centered on the QTP, which we term the \textbf{QTP-Mode}.

A defining feature of the QTP-Mode is its tripolar architecture, linking the QTP with the Arctic and Antarctica through robust atmospheric corridors. These connections are not transient statistical associations, but are sustained by identifiable physical pathways. Variability associated with the Arctic Oscillation and Antarctic Oscillation, together with polar ice-driven circulation changes, modulates coupling strength and governs the propagation of Rossby wave activity across hemispheres. The necessity of both mechanical and thermal forcing for maintaining this structure is confirmed by targeted numerical experiments in which removing either QTP topography or vertical diffusive heating substantially weakens large-scale teleconnections while enhancing local interactions. This result establishes that the QTP-Mode emerges from the combined action of topographic dynamics and diabatic heating, rather than from either mechanism alone.

Beyond revealing a tripolar structure, our findings indicate that long-range climate connectivity is not spatially diffuse, but increasingly organized around dynamically sensitive regions. Under ongoing warming, teleconnection intensity becomes more concentrated along evolving snow and sea-ice boundaries, particularly at polar margins. These interfaces act not merely as passive boundaries, but as active zones where thermodynamic gradients, storm tracks, and waveguides interact to shape planetary-scale coupling. Within this framework, the QTP emerges not simply as a downstream recipient of remote variability, but also as an active transmitter of influence, capable of modulating the structure and stability of hemispheric-scale feedback systems. This reinterpretation challenges prevailing views that treat high-elevation regions as peripheral actors in global climate dynamics and instead positions the QTP as a key organizing element in the Earth system.

From a methodological perspective, the present study highlights the power and flexibility of the CN framework for uncovering emergent structures in the climate system. The robustness of the QTP-Mode across multiple thresholds, variables, and model ensembles underscores that the identified teleconnections reflect physically embedded dynamics rather than statistical artifacts. Compared with trajectory-based transport models such as HYSPLIT, which require detailed atmospheric state variables and are primarily retrospective, the CN approach extracts global interaction patterns using only near-surface temperature. This minimal data requirement enables systematic exploration across historical observations and future projections, and allows natural extension to temporal~\cite{LiFundamental2017,HolmeTemporal2012}, multilayer~\cite{DeDomenicoMore2023,BoccalettiStructure2014}, and higher-order network frameworks~\cite{MillanTopology2025,RosasDisentangling2022}. Such approaches are increasingly essential for capturing nonlinear interactions, feedback amplification, and cascading behavior among climate tipping elements.

Physically, our findings indicate that the three poles, the QTP, Arctic and Antarctic, form an active and dynamically coupled system that transmits climate signals across hemispheres, rather than acting as isolated or purely passive components of the climate system. The amplified warming of the QTP, together with accelerating Arctic and Antarctic ice loss, implicates cryospheric feedbacks in the active reorganization of planetary teleconnections. The spatial alignment between shifting sea-ice boundaries and changes in the tripolar teleconnection pattern supports this interpretation. While uncertainties remain in the magnitude and regional expression of these feedbacks, particularly across models, our results reinforce a growing consensus that Earth system stability depends not only on the behavior of individual tipping elements, but critically on the structure of their interactions~\cite{BoersDestabilization2025}.

More broadly, this study advances a systems-level perspective on climate risk. 
By mapping the QTP’s embeddedness within a planetary-scale interaction network linking polar regions, tropical systems, and major climate tipping elements, our results highlight how highland–lowland and pole-to-pole feedbacks may reorganize under continued warming. As the likelihood of cascading tipping events increases~\cite{WunderlingGlobal2023}, a central challenge for future research will be to identify, quantify, and monitor these corridors of influence in a dynamic manner. Embedding network-aware diagnostics into Earth system models and early-warning frameworks offers a promising pathway toward anticipating systemic transitions in an increasingly interconnected climate system~\cite{HelbingGlobally2013}.

\clearpage

\clearpage

\begin{figure}
\begin{centering}
  \vspace{-12mm}
  \includegraphics[width=1\textwidth]{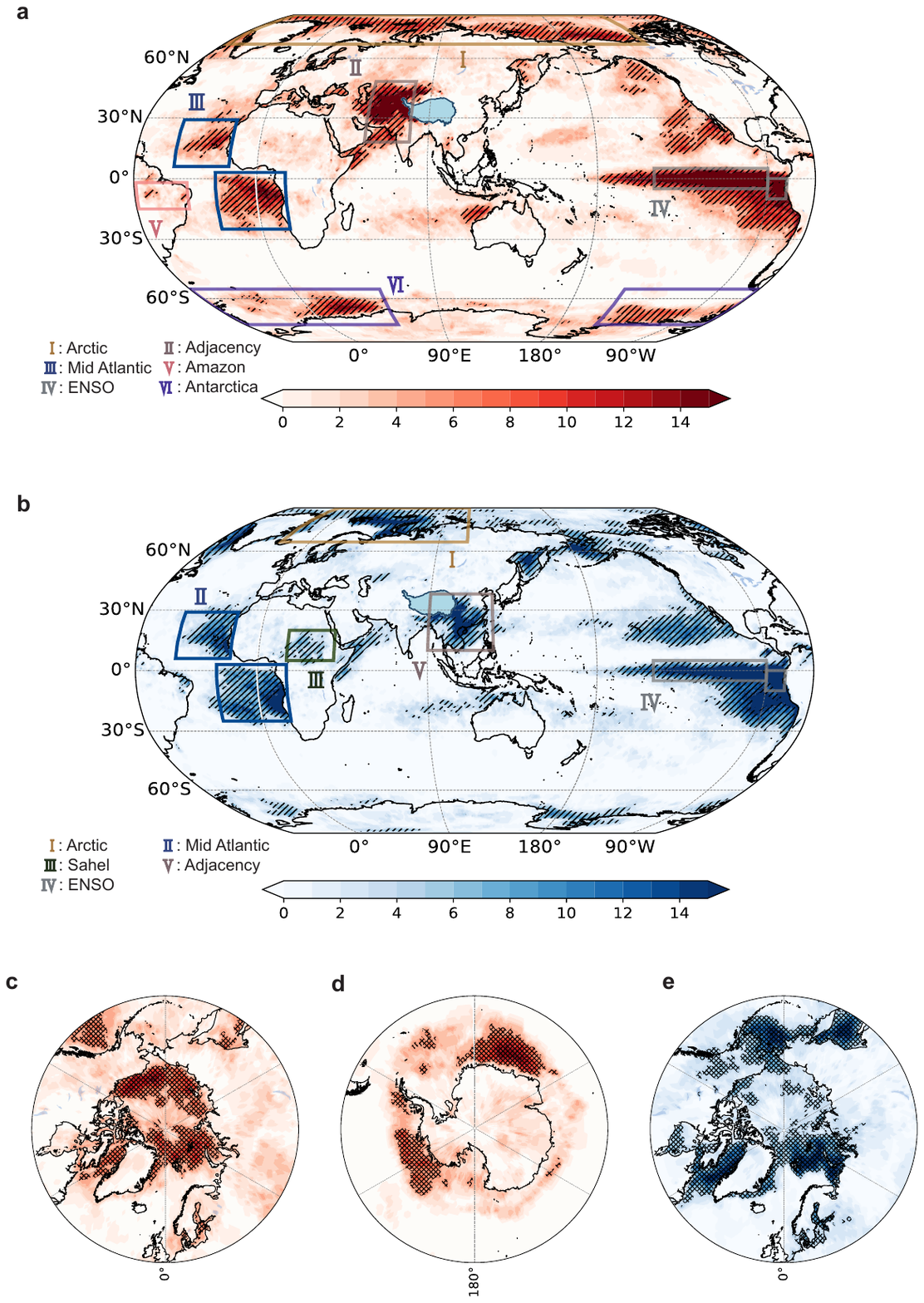}
  \par\end{centering}
  \captionsetup{labelformat=empty}
  \caption{\label{Fig2:pattern}}
  \label{Fig2}
\end{figure}
\clearpage
\noindent\textbf{Fig.~2. The QTP-Mode and its planetary-scale teleconnection structure.} Values at each grid point represent the frequency with which statistically significant, directional connections with the QTP are detected over the 43-year analysis period, with significance assessed against a shuffle-based null model. Higher values indicate more persistent and robust climate interactions. \textbf{a}, Global distribution of $F_{\text{out}}$, quantifying the directional influence of external regions on the QTP. Prominent and stable influences emerge from the Arctic, Adjacency, tropical Atlantic, Amazon, ENSO, and Antarctica (boxed regions I–VI), indicating coherent large-scale forcing pathways toward the Plateau. \textbf{b}, Distribution of $F_{\text{in}}$, highlighting regions that are significantly influenced by the QTP, including the Arctic, tropical Atlantic, Sahel, Adjacency and ENSO domain (boxed regions I–IV), revealing the outward reach of Plateau-driven interactions. \textbf{c–e}, Polar projections illustrating the spatial organization and directionality of the tripolar teleconnection structure. Panels \textbf{c} and \textbf{d} show the outward influence field on the QTP from the Arctic and Antarctica respectively, while \textbf{e} illustrates the feedback influence of the QTP on Arctic regions.
\clearpage

\begin{figure}
  \centering
  \vspace{-12mm}
  \includegraphics[width=0.9\textwidth]{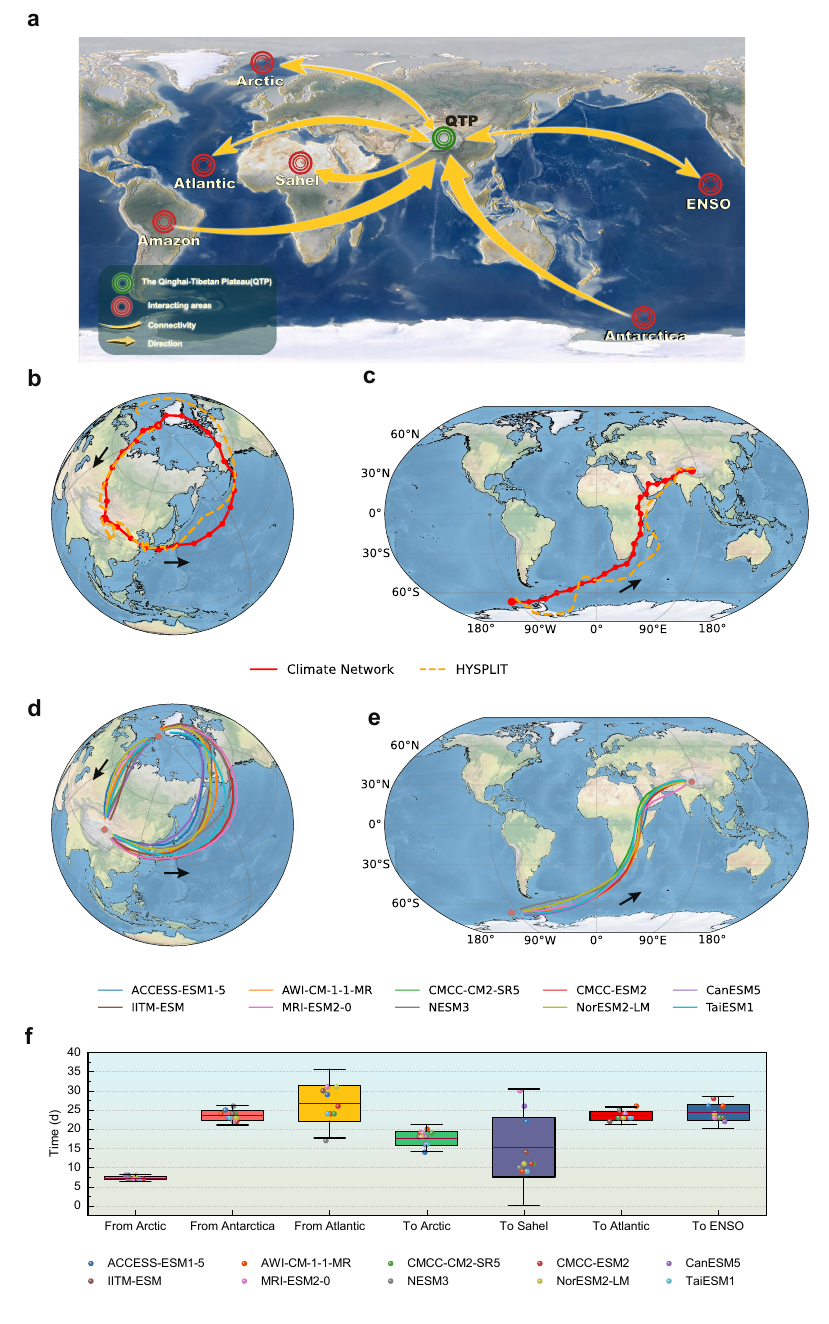}
  \captionsetup{labelformat=empty}
  \caption{\label{Fig3:paths}}
\end{figure}
\clearpage
\noindent\textbf{Fig.~3. Propagation pathways and physical mechanisms of QTP teleconnections.} \textbf{a}, Conceptual overview of the QTP-Mode, illustrating how the QTP is embedded within a planetary-scale teleconnection structure linking major climate tipping elements. \textbf{b, c}, Dominant propagation pathways connecting the QTP with the Arctic and Antarctica, respectively. Arrows indicate the directions. Red curves show pathways inferred from the CN analysis, while orange curves denote atmospheric transport trajectories simulated using the HYSPLIT model, demonstrating close spatial agreement between statistical inference and physically resolved transport. \textbf{d, e}, Robust historical propagation pathways reconstructed from CMIP6 multi-model simulations, revealing a consistent spatial organization of QTP-related teleconnections across models. \textbf{f}, Distribution of propagation times for seven representative teleconnection pathways derived from observational and model datasets. Propagation times are computed by summing lag intervals between successive nodes along each pathway. Box plots indicate stable and comparable transit times across pathways and datasets.
\clearpage

\begin{figure}
\begin{centering}
  \includegraphics[width=0.9\textwidth]{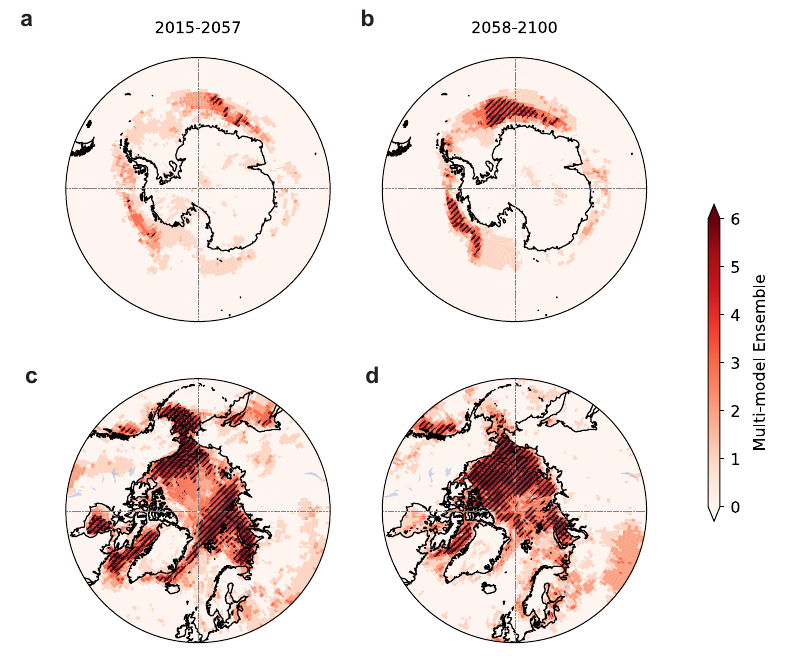}
  \par\end{centering}
  \captionsetup{labelformat=empty}
  \caption{\label{Fig4:FuturePattern}}
\end{figure} 
\noindent\textbf{Fig.~4. Projected evolution of QTP-related tripolar teleconnections under climate change.} Multi-model ensemble representations of the QTP-Mode derived from CMIP6 simulations under the high-emission scenario SSP5–8.5 are shown for two future 43-year periods: the early future (2015–2057; \textbf{a, c}) and the late future (2058–2100; \textbf{b, d}). \textbf{a, b}, Ensemble-mean distribution of $F_{\text{out}}$, quantifying the directional influence of external regions on the QTP under future warming. \textbf{c, d}, Ensemble-mean distribution of $F_{\text{in}}$, highlighting regions that are increasingly influenced by the QTP. Color shading indicates the number of CMIP6 models (out of ten) exhibiting statistically significant and persistent teleconnections with the QTP at each grid point, based on thresholds applied consistently across models and time periods.

\section*{Methods}
\subsection*{Data}
{\noindent} We employ three complementary datasets to quantify, validate, and project climate teleconnections associated with the QTP.

First, near-surface (2 m) air temperature was obtained from the ERA5 global reanalysis~[71], 
produced by the European Centre for Medium-Range Weather Forecasts, covering January 1979 to December 2021. Near-surface temperature provides a physically interpretable and widely used measure of large-scale climate variability and teleconnections~[60, 72].
Daily values at 00:00 UTC were extracted to represent synoptic-scale thermal conditions. To ensure computational efficiency while preserving large-scale structure, the original $0.25^\circ \times 0.25^\circ$ grid was downsampled to $1^\circ \times 1^\circ$ by retaining one grid point in each $4 \times 4$ block, yielding a global grid of 181 × 360 nodes. Leap days were removed to maintain temporal consistency. Seasonal variability associated with the annual cycle was removed by subtracting the long-term mean daily climatology at each grid point, isolating anomalous temperature fluctuations. To assess variable sensitivity and ensure methodological robustness, the same analysis was repeated using ERA5 air temperature at 1000 hPa.

Second, to examine the robustness of the QTP-Mode across different climate states and future forcing scenarios, we analyzed outputs from multiple Coupled Model Intercomparison Project Phase 6 (CMIP6) models. Historical simulations span 1950–2014, and future projections follow the high-emission Shared Socioeconomic Pathway SSP5–8.5 for 2015–2100. Depending on model availability, near-surface air temperature (\textit{tas}) and atmospheric temperature at selected pressure levels (\textit{ta}) were used. A complete list of models and associated metadata is provided in Supplementary Table~S1.

Third, to independently evaluate the physical plausibility of network-inferred teleconnection pathways, we employed the Hybrid Single-Particle Lagrangian Integrated Trajectory (HYSPLIT) model. Meteorological forcing was taken from the Global Data Assimilation System (GDAS) archive provided by the National Centers for Environmental Prediction (NCEP), which supplies three-dimensional wind, temperature, and humidity fields at $1^\circ \times 1^\circ$ spatial resolution and 3-hourly temporal resolution. This dataset is well suited for simulating long-range atmospheric transport relevant to QTP–polar and intercontinental teleconnections.

\subsection*{Climate network construction}
{\noindent}To quantify the global influence structure of the QTP, we constructed directed CNs based on time-lagged correlations between the QTP and the rest of the global climate system. Network nodes were partitioned into two communities according to geographic boundaries: $N_1 = 252$ nodes located within the QTP region and $N_2 = 64,908$ nodes outside the QTP. This configuration enables explicit assessment of directional interactions between the Plateau and external regions.

Using ERA5 temperature anomalies, we constructed one CN for each year from 1979 to 2021. For a given year $y$, we computed the time-lagged cross-correlation function between temperature anomaly time series $S_i(d)$ and $S_j(d)$, where $i$ denotes a QTP node and $j$ denotes a non-QTP node. Here, $d$ represents the start day of the year, and each series has a length of 365 days. Correlations were evaluated as,
\begin{equation}
C_{i,j}^y(\tau) = \frac{\langle S_i(d) S_j(d+\tau) \rangle - \langle S_i(d) \rangle \langle S_j(d+\tau) \rangle}{\sqrt{\langle (S_i(d) - \langle S_i(d) \rangle)^2 \rangle} \cdot \sqrt{\langle (S_j(d+\tau) - \langle S_j(d+\tau) \rangle)^2 \rangle}},
\label{EQ:C1}
\end{equation}
and
\begin{equation}
C_{i,j}^y(-\tau) = \frac{\langle S_i(d+\tau) S_j(d) \rangle - \langle S_i(d+\tau) \rangle \langle S_j(d) \rangle}{\sqrt{\langle (S_i(d+\tau) - \langle S_i(d+\tau) \rangle)^2 \rangle} \cdot \sqrt{\langle (S_j(d) - \langle S_j(d) \rangle)^2 \rangle}},
\label{EQ:C2}
\end{equation}
where $\langle \cdot \rangle$ denotes temporal averaging within the year.

We set the maximum lag to $\tau_{\max}=200$ days~[28,73]
, yielding 401 lag values per node pair. For each pair, the edge weight was defined as the signed correlation coefficient $C^y(\tau_*)$ corresponding to the lag $\tau_*$ that maximizes the absolute correlation. The sign and magnitude of $C^y(\tau_*)$ determine the strength of interaction, while the sign of $\tau_*$ determines directionality. A positive $\tau_*$ indicates that the QTP node leads the external node, corresponding to an outward (OUT) edge from the QTP. A negative $\tau_*$ indicates an inward (IN) edge toward the QTP. Zero-lag correlations ($\tau_*=0$) were excluded because they do not provide a clear causal ordering.

The directed adjacency matrices for year $y$ are defined as,
\begin{equation}
\mathrm{A}_{i,j}^y = (1-\delta_{i,j}) H(\tau_*),
\end{equation}
\begin{equation}
\mathrm{A}_{j,i}^y = (1-\delta_{i,j}) H(-\tau_*),
\end{equation}
where $\delta_{i,j}$ is the Kronecker delta and $H(x)$ is the Heaviside step function.

To retain only statistically meaningful interactions and suppress spurious correlations, we applied a dynamic, year-specific significance threshold. For each CN, only correlations exceeding the 95th percentile of positive values or below the 5th percentile of negative values were retained. Denoting these thresholds as $C^y_{\mathrm{in},+}$, $C^y_{\mathrm{in},-}$, $C^y_{\mathrm{out},+}$, and $C^y_{\mathrm{out},-}$, Fig.~S1d illustrates the probability density function (PDF) of edge weights $C^y_{ij,+}$ in the climate networks constructed for $y = 1979$, $2000$, and $2020$ as examples, and the final adjacency matrices are given by,
\begin{equation}
\mathrm{X}_{i,j}^y = \mathrm{A}_{i,j}^y \cdot H[(C^y(\tau_*) - C^y_{\mathrm{in},+})(C^y(\tau_*) - C^y_{\mathrm{in},-})],
\label{EQ:QTP-OUT}
\end{equation}
and
\begin{equation}
\mathrm{X}_{j,i}^y = \mathrm{A}_{j,i}^y \cdot H[(C^y(\tau_*) - C^y_{\mathrm{out},+})(C^y(\tau_*) - C^y_{\mathrm{out},-})].
\label{EQ:IN-QTP}
\end{equation}

Applying this procedure across 43 years yields 84 directed networks accounting for both outward and inward connections in each year. The number is not 86 networks because the data from the last year was used for lag calculation. This framework enables systematic, year-by-year analysises of the directionality, persistence, and evolution of teleconnections linking the QTP to the global climate system.

\subsection*{Network statistics}
{\noindent}Following network construction, we quantified directional climate interactions between the QTP and the rest of the globe using node degree statistics. Degree measures provide a compact representation of how strongly and persistently each external region is dynamically linked to the QTP.

For each external node $j$ and year $y$, the out-degree $D_{\mathrm{out}}^y(j)$ is defined as the number of statistically significant connections from node $j$ to nodes within the QTP,
\begin{align}
    D_{\mathrm{out}}^y(j) &= \sum_{i \in \text{QTP}} {X}_{j,i}^{y},
    \label{EQ:outdegree}
\end{align}
quantifying the strength of influence exerted by external regions on the QTP. Conversely, the in-degree $D_{\mathrm{in}}^y$ measures the number of significant links from the QTP to node $j$,
\begin{align}
    D_{\mathrm{in}}^y(j) &= \sum_{i \in \text{QTP}} {X}_{i,j}^{y},
    \label{EQ:indegree}
\end{align}
representing the influence of the QTP.

The spatial distributions of $D_{\mathrm{out}}^y(j)$ and $D_{\mathrm{in}}^y(j)$ therefore characterize the bidirectional structure of climate interactions between the Plateau and each global location. Larger degree values indicate more frequent and robust teleconnections.

Degree distributions exhibit a highly skewed structure (Fig.~S1e), with most nodes displaying weak or intermittent connections and a small subset showing persistent high-intensity interactions. To isolate these structurally dominant nodes, we applied an annually adaptive threshold $D_{*,\mathrm{out}}^{y}$ and $D_{*,\mathrm{in}}^{y}$, defined as the 95th percentile of the degree distribution in each year. Binary indicators of significant influence are then defined as,
\begin{equation}
P_{\mathrm{out}}^y(j) = 
\begin{cases}
1, & D_{\mathrm{out}}^y(j) > D_{*,\mathrm{out}}^{y}\\
0, & D_{\mathrm{out}}^y(j) \leq D_{*,\mathrm{out}}^{y},
\end{cases}
\end{equation}
and
\begin{equation}
P_{\mathrm{in}}^y(j) = 
\begin{cases}
1, & D_{\mathrm{in}}^y(j) > D_{*,\mathrm{in}}^{y}\\
0, & D_{\mathrm{in}}^y(j) \leq D_{*,\mathrm{in}}^{y}.
\end{cases}
\end{equation}

To quantify the temporal persistence of strong interactions, we computed the cumulative frequency with which each node appears among the top 5\% of degree values across the 43-year period. The resulting metrics,
\begin{equation}
    F_{\mathrm{out}} = \sum_{y} P_{\mathrm{out}}^y,
    \label{EQ:F_OUT}
\end{equation}
and
\begin{equation}
    F_{\mathrm{in}} = \sum_{y} P_{\mathrm{in}}^y,
    \label{EQ:F_IN}
\end{equation}
measure the long-term stability of outward and inward teleconnections associated with the QTP, respectively.

Statistical significance was assessed using a surrogate-based null model. Temperature time series were temporally shuffled to destroy physical causality while preserving marginal distributions, and the full CN construction and degree analysis were repeated for each surrogate realization. This procedure yields null distributions of $F_{\mathrm{out}}^{\mathrm{shuffled}}$ and $F_{\mathrm{in}}^{\mathrm{shuffled}}$. Nodes with observed frequencies exceeding the 90th percentile of the corresponding null distribution were deemed statistically significant, as can be seen in Extend Fig.~1, and are indicated by crosses in Fig.~\ref{Fig2:pattern}.

\subsection*{Causal Inference Frameworks}

The QTP-Mode was initially identified through correlation analysis. However, correlation alone is insufficient to distinguish directional causality from mere statistical association, nor can it rule out spurious links induced by common drivers. To rigorously verify the physical robustness and directional influence of the QTP-Mode, we employ two distinct causal inference frameworks: Granger Causality analysis and the Liang Information Flow method.

Prior to the causality analysis, we performed spatial averaging on the date series of nodes that exhibited significant F-values to construct representative time series. All resulting time series were subjected to the Augmented Dickey-Fuller (ADF) test to ensure stationarity, a prerequisite for the following causal evaluations.

\subsubsection*{Granger Causality Analysis}
We first employ the Granger Causality (GC) test to assess the predictive relationship between the QTP-Mode and the target climate fields. Following the methodology described by Silva et al. (2021), GC is based on the principle that if a variable $Y$ causes $X$, then past values of $Y$ should contain information that helps predict $X$ beyond what is contained in the past values of $X$ alone.

For two time series $X(t)$ and $Y(t)$, we consider two linear autoregressive models. The \textit{restricted model} (null hypothesis) assumes $X(t)$ depends only on its own past values:
\begin{equation}
X(t) = \gamma_0 + \sum_{\tau=1}^{\tau_{max}} \gamma_{\tau} X(t-\tau) + \epsilon_r(t)
\end{equation}
The \textit{complete model} assumes $X(t)$ depends on the past values of both $X$ and $Y$:
\begin{equation}
X(t) = \alpha_0 + \sum_{\tau=1}^{\tau_{max}} \alpha_{\tau} X(t-\tau) + \sum_{\tau=1}^{\tau_{max}} \beta_{\tau} Y(t-\tau) + \epsilon_c(t)
\end{equation}
where $\tau_{max}$ represents the maximum time lag, and $\epsilon_r, \epsilon_c$ are the residuals. 

We say that $Y$ Granger-causes $X$ if the inclusion of $Y$ significantly reduces the prediction error. This is quantified by comparing the residual sum of squares (RSS) of the two models using an F-test or a Chi-square approximation (Hamilton, 1994; Silva et al., 2021):
\begin{equation}
F = \frac{(RSS_r - RSS_c) / \tau_{max}}{RSS_c / (N - 2\tau_{max} - 1)}
\end{equation}
A significant F-statistic implies that the history of $Y$ exerts a statistically significant influence on the current state of $X$.

\subsubsection*{Liang Information Flow Method}
While Granger causality relies on statistical predictability, the Information Flow (IF) theory proposed by Liang (2014) provides a rigorous measure of causality derived from the first principles of dynamical systems. This method quantifies the rate of information transfer between time series without requiring the manual selection of lag parameters, which is often a challenge in traditional lag-correlation or GC analysis.

For a linear stochastic system, the rate of information flow from $Y$ to $X$ (denoted as $T_{Y \to X}$) can be estimated using the maximum likelihood estimator (MLE):
\begin{equation}
T_{Y \to X} = \frac{C_{XX} C_{XY} C_{Y, dX} - C_{XY}^2 C_{X, dX}}{C_{XX}^2 C_{YY} - C_{XX} C_{XY}^2}
\end{equation}
where $C_{ij}$ denotes the sample covariance between time series $i$ and $j$, and $d X$ represents the time difference approximation of $X$ (Euler forward differencing). 

The value of $T_{Y \to X}$ offers a clear physical interpretation:
\begin{itemize}
    \item $T_{Y \to X} = 0$: No causal link exists from $Y$ to $X$.
    \item $T_{Y \to X} \neq 0$: A causal link exists. Specifically, a positive value indicates that $Y$ functions to destabilize $X$, while a negative value indicates a stabilizing effect.
\end{itemize}

By combining the statistical predictability of Granger Causality with the dynamical information transfer of Liang's method, we ensure that the identified QTP-Mode teleconnections are both statistically significant and physically grounded.

\subsection*{Teleconnection pathway identification}
{\noindent}To identify the most efficient teleconnection pathways embedded within the CN, we employed a shortest-path framework grounded in network theory. A high-resolution directed CN was constructed using 6,570 quasi-uniformly distributed nodes covering the global domain~\cite{mengPercolationFrameworkDescribe2017}, without any regional partitioning. 

For each year $y$, the weight assigned to an edge connecting nodes $i$ and $j$ was defined using the standardized anomaly of the peak time-lagged correlation~[22,75]
,
\begin{equation}
  W_{i,j}^y = \frac{C_{i,j}^y(\tau_*) - \mathrm{mean}(C_{i,j}^y(\tau))}{\mathrm{std}(C_{i,j}^y(\tau))},
  \label{Weight}
\end{equation}
where $C_{i,j}^y(\tau)$ denotes the time-lagged cross-correlation function computed from Eqs.~(\ref{EQ:C1}) and (\ref{EQ:C2}), and $\tau_*$ is the lag at which the absolute correlation reaches its maximum. This normalization ensures that edges associated with unusually strong correlation peaks relative to the local lag structure receive higher weights, independent of absolute variance.

To translate statistical association into a traversal metric, we defined the cost function of each edge as the inverse magnitude of its weight,
\begin{equation}
\text{CF}_{i,j}^y = \frac{1}{|W_{i,j}^y|}.
  \label{cost_function}
\end{equation}

Under this definition, optimal paths preferentially follow links with stronger and more statistically distinct correlations, thereby enhancing the physical plausibility of inferred propagation routes~[13].
Directed optimal paths between selected node pairs were computed using Dijkstra’s algorithm~\cite{Dijkstra1959}, subject to two geophysical constraints. First, the great-circle distance between consecutive nodes was restricted to be less than 1,000 km, ensuring spatial continuity consistent with synoptic-scale atmospheric processes~[47]
. Second, the direction of each step was required to be consistent with the sign of the optimal lag $\tau_*$, thereby preserving the inferred direction of information flow and excluding non-causal transitions.

\subsection*{Lagrangian validation of teleconnection pathways}
\noindent To assess the physical plausibility of the teleconnection pathways inferred from the CN analysis, we employed the HYSPLIT model~[54] 
to simulate three-dimensional atmospheric parcel transport, with a focus on large-scale water-vapor–relevant airflow.

HYSPLIT is a well-established atmospheric transport and dispersion model developed by the Air Resources Laboratory of the National Oceanic and Atmospheric Administration in collaboration with the Australian Bureau of Meteorology. It combines Lagrangian and Eulerian formulations to resolve air-parcel trajectories under the influence of advection, diffusion, and vertical mixing, and is widely used to investigate synoptic to intercontinental transport processes. In this study, we employed the trajectory ensemble configuration to account for uncertainty in atmospheric flow, generating multiple trajectories from each source location by perturbing initial conditions in the horizontal and vertical directions. Following the default HYSPLIT setup, vertical perturbations of approximately 0.01 sigma units (roughly 250m) were applied, enabling realistic representation of free-tropospheric mixing.

To ensure consistency with the propagation time scales inferred from the CN analysis, each trajectory was integrated for a maximum duration of 30 days. Given the elevated topography and deep boundary-layer structure of the QTP, trajectories were initialized at an altitude of 5,000m above sea level. Meteorological forcing fields were obtained from the GDAS, which provides global wind, temperature, and humidity fields at $1^\circ \times 1^\circ$ spatial resolution and 3-hourly temporal resolution.

Both forward and backward trajectories were computed from selected grid points within the QTP to characterize outgoing and incoming atmospheric transport pathways. These Lagrangian trajectories were then compared with the network-derived teleconnection paths to provide independent physical validation and mechanistic interpretation. While HYSPLIT ensemble simulations generate multiple closely spaced trajectories owing to intrinsic atmospheric variability, the pathway shown is representative rather than selectively chosen; the ensemble exhibits robust convergence in large-scale structure, and our interpretation is insensitive to the specific trajectory displayed. For instance, the backward trajectory representing Arctic-to-QTP influence (Fig.~\ref{Fig2:pattern}b) was initialized in January 2014 from 30.0°N, 84.0°E, corresponding to wintertime Arctic forcing. The forward trajectory illustrating QTP-to-Arctic propagation (Fig.~\ref{Fig2:pattern}e) was initiated in July 2022 from 37.0°N, 93.0°E, consistent with summertime downstream influence from the Plateau. A representative Antarctica-to-QTP pathway (Fig.~\ref{Fig2:pattern}c) was initialized in July 2020 from 34.0°N, 88.0°E.

\subsection*{Dynamic and thermodynamic mechanisms of the QTP-Mode}

\noindent To disentangle the respective roles of dynamical and thermodynamical processes in shaping the QTP-Mode, we combined the CN framework with targeted numerical experiments using a fully coupled Earth system model.

\textbf{Model configuration and experimental design}: Simulations were conducted with the Flexible Global Ocean–Atmosphere–Terrestrial System model, finite-volume version 3 (FGOALS-f3-L)~\cite{HeCAS2020,Xu2025Model}. This model has been specifically employed to investigate the climatic effects of Asian orography through a series of sensitivity experiments under CMIP6 AMIP protocols. To probe the underlying QTP-mode mechanisms, we analyze a set of the latest sensitivity experiments, including control, dynamic perturbation, and thermodynamic perturbation runs. The perturbations are applied to the QTP and the adjacent Iranian Plateau as a combined domain. This constitutes the most targeted available model configuration for isolating the mechanical and thermal forcing of the QTP region. The specific configurations of each experiment are as follows:

\textit{Control run (amip)}: The control simulation retains realistic topography and surface energy exchange over the QTP and surrounding regions, serving as the baseline for comparison.

\textit{No mountain run (amip\underline{~~}NMO)}: To suppress mechanical forcing, surface elevations exceeding 500m over the QTP and adjacent regions were reduced to 500m, thereby removing orographic lifting and terrain-induced circulation effects while preserving surface thermal processes.

\textit{No sensible heating run (amip\underline{~~}NS\underline{~~}MO)}: To suppress thermodynamical forcing, the original topography was retained, but the vertical diffusive heating term associated with surface sensible heat flux was set to zero over grid cells above 500m, effectively eliminating surface-driven diabatic heating while maintaining mechanical forcing.

The model integration starts on January 1, 1976, with the external forcing factors, including greenhouse gasses, solar irradiance, ozone, SSTs, and sea ice, as defined by their historical and observed values. The first three integration years are recognized as the spin-up time. Here, we utilize the latest model output, specifically the daily 2 m near-surface temperature data covering 1979–2020. Fig.~S12 shows the 2 m temperature differences among these three experiments: the amip\underline{~~}NMO experiment exhibits higher temperatures than amip due to weakened orographic lifting, while the amip\underline{~~}NS\underline{~~}MO experiment shows lower temperatures than amip due to suppressed vertical thermal diffusion (taking August 1, 2020, as an example).

\textbf{CN analysis}: CNs were constructed using the same time-lagged Pearson correlation methodology applied to observational data, yielding teleconnection frequency metrics $F_{\mathrm{amip}}$, $F_{\mathrm{amip\underline{~~}NMO}}$, and $F_{\mathrm{amip\underline{~~}NS\underline{~~}MO}}$.

To quantify the contributions of dynamical and thermodynamical processes, difference fields were computed as dynamic effect,
\begin{equation}
    \Delta F_{\mathrm{dynamic}} = F_{\mathrm{amip\underline{~~}NMO}} - F_{\mathrm{amip}},
    \label{EQ:DynamicEffect}
\end{equation}
and thermal effect,
\begin{equation}
    \Delta F_{\mathrm{thermal}} = F_{\mathrm{amip\underline{~~}NS\underline{~~}MO}} - F_{\mathrm{amip}}.
    \label{EQ:ThermalEffect}
\end{equation}

\textit{Significance testing}: 
Significance of the teleconnection changes was assessed using a shuffled null model consistent with the observational network analysis. For each experiment, temporally shuffled temperature fields were used to generate $F_{\mathrm{amip}}^{\mathrm{shuffled}}$, $F_{\mathrm{amip\underline{~~}NMO}}^{\mathrm{shuffled}}$, and $F_{\mathrm{amip\underline{~~}NS\underline{~~}MO}}^{\mathrm{shuffled}}$. Under the null hypothesis that perturbation-induced differences arise from random variability, difference distributions were constructed as
\begin{equation}
\Delta F_{\mathrm{Dynamic}}^{\mathrm{null}} = F_{\mathrm{amip\underline{~~}NMO}}^{\mathrm{shuffled}} - F_{\mathrm{amip}}^{\mathrm{shuffled}},
\end{equation}
and
\begin{equation}
\Delta F_{\mathrm{Thermal}}^{\mathrm{null}} = F_{\mathrm{amip\underline{~~}NS\underline{~~}MO}}^{\mathrm{shuffled}} - F_{\mathrm{amip}}^{\mathrm{shuffled}}.
\end{equation}

Grid-point differences exceeding the 95th percentile or below the 5th percentile of the corresponding null distributions were considered statistically significant.

\textbf{Regional diagnostics}.  
To quantify changes over key components of the QTP-Mode, regional statistics were computed for predefined domains, including the Arctic, Antarctica, ENSO region, and others. Two metrics were evaluated: (1) the mean absolute change in significant $\Delta F$ values within each region, and (2) the relative change rate, defined as the ratio of absolute change to the regional mean $F_{\mathrm{amip}}$. Results are shown in Table.~S3.

\subsection*{Multi-model assessment}
{\noindent} To assess the evolution of QTP-Mode teleconnections under projected climate warming, we applied the CN framework to simulations from ten CMIP6 models (Table~S1). Analyses were conducted for two future periods, 2015–2057 and 2058–2100, under the high-emission SSP5–8.5 scenario. For each model $m$ and time window, CNs were constructed and QTP-Modes were identified using the same procedures applied to the observational analysis, yielding a total of 20 model-based QTP-Modes across all models and periods (Fig.~S17).

To synthesize information across models, we computed the multi-model frequency of statistically significant teleconnections between each grid point and the QTP. These ensemble measures are denoted as $FM_{\mathrm{out}}$ and $FM_{\mathrm{in}}$ and defined as,
\begin{equation}
    FM_{\mathrm{out}} = \sum_{m} F_{\mathrm{out}}^m,
    \label{EQ:FModel_OUT}
\end{equation}
and
\begin{equation}
    FM_{\mathrm{in}} = \sum_{m} F_{\mathrm{in}}^m,
    \label{EQ:FModel_IN}
\end{equation}
where $F_{\mathrm{out}}^m$ and $F_{\mathrm{in}}^m$ denote the model-specific frequency maps for the outward and inward directions, respectively. The maximum possible value at each grid point is 10, corresponding to agreement across all models in the ensemble.

Statistical robustness was evaluated using a shuffling-based null model applied separately to each future period. For each grid point, null distributions of $FM_{\mathrm{out}}$ and $FM_{\mathrm{in}}$ were constructed from randomized versions of each model’s time series. Grid points exceeding the 90th percentile of the corresponding null distribution were classified as statistically significant, indicating teleconnection patterns that are consistent across models and unlikely to arise from random variability.

\section*{Data availability}
The ERA5 reanalysis data used here are publicly available at: \url{https://cds.climate.copernicus.eu/cdsapp#!/dataset/reanalysis-era5-single-levels}.
The CMIP6 data are publicly available at: \url{https://esgf-node.llnl.gov/projects/cmip6/}.
The Arctic sea ice concentration product is available at:
\url{https://cds.climate.copernicus.eu/cdsapp#!/dataset/satellite-sea-ice-concentration?tab=overview}. The archived data used for the HYSPLIT model are available at : \url{https://www.ready.noaa.gov/data/archives/gdas1/}. The newest daily mean T2m of model output is available in zenodo: \url{https://doi.org/10.5281/zenodo.18091229}.
\section*{Code availability}
The Python codes used for the analysis are available on GitHub (\url{https://github.com/jingfang/QTP}).

\section*{Acknowledgments}
The authors acknowledge the support of the National Natural Science Foundation of China (Grant No. T2525011, 42450183, 12275020, 12135003, 12205025, 42461144209), the National Key Research and Development Program of China (grant no. 2025YFF0517304) and the Fundamental Research Funds for the Central Universities.

{\section*{Author Contributions}
J.F. designed the research. Z.W. performed the analysis, and all authors discussed the results and contributed to writing the manuscript. J.F. led the writing of the manuscript.}

\section*{Additional information}
Supplementary Information is available in the online version of the paper.

\section*{Competing interests}
The authors declare no competing interests.

\clearpage
\begin{appendices}
\setcounter{table}{0}
\renewcommand{\thetable}{S\arabic{table}}%
\setcounter{figure}{0}
\renewcommand{\thefigure}{Extended Data~\arabic{figure}}%


\begin{figure}[htbp]
  \centering
  \includegraphics[width=1.\textwidth]{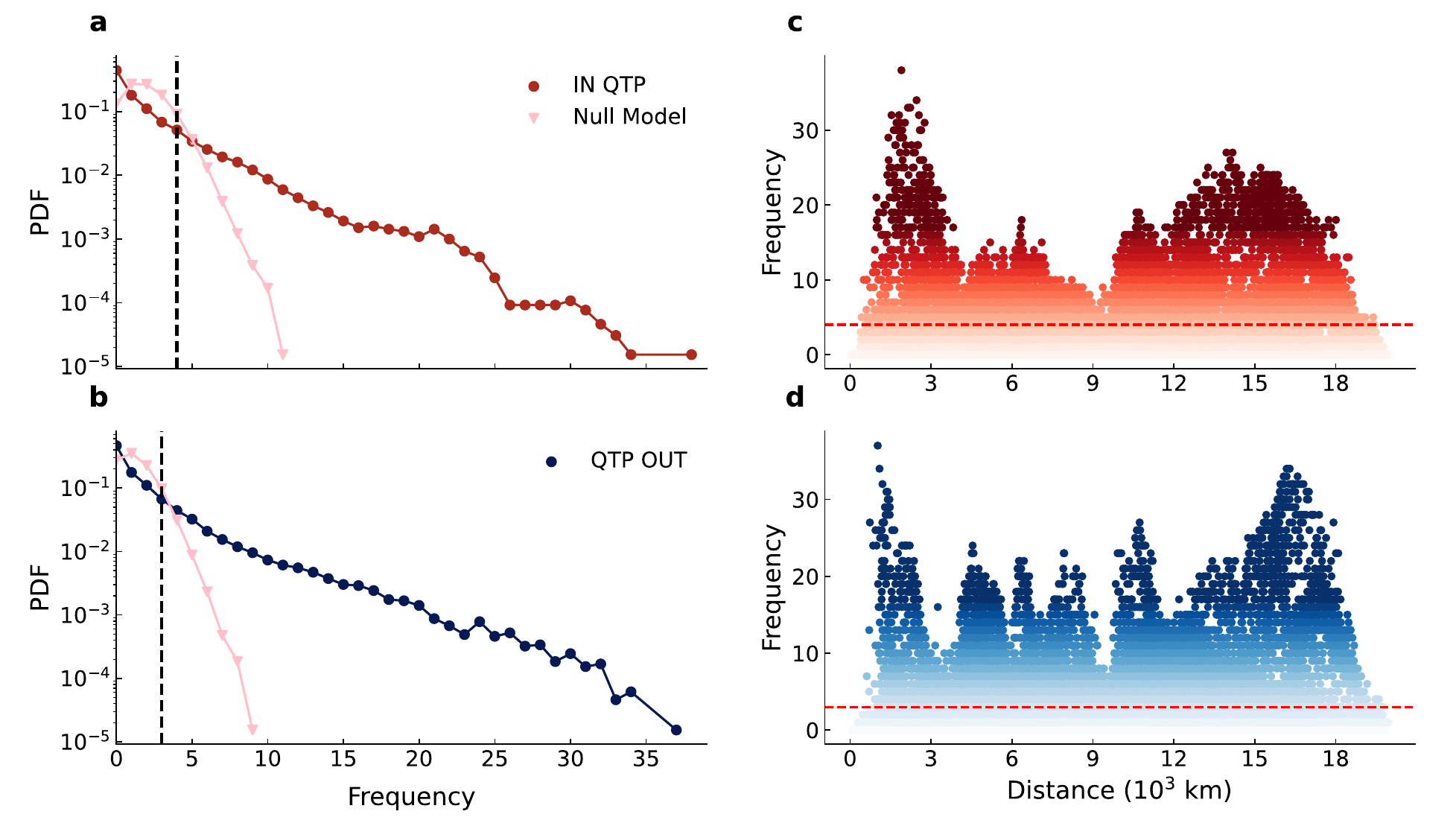}
  \captionsetup{labelformat=empty}
  \caption{\label{SFig_Q3}}
\end{figure} 
\noindent\textbf{Extended Data Fig.~1~\(|\) Statistical characteristics of QTP-related teleconnections.} \textbf{a, b}, Frequency distributions of nodes exhibiting high connectivity with the QTP over the 1979–2021 period, quantified by \( F_{\text{out}} \) and \( F_{\text{in}} \), which measure the persistence of outward and inward teleconnections, respectively. Both distributions display approximately exponential decay and show systematically longer tails than those obtained from shuffled null models, indicating enhanced persistence of QTP-related connections. \textbf{c, d}, Dependence of \( F_{\text{out}} \) and \( F_{\text{in}} \) on geographic distance from the QTP. Pronounced peaks are evident at distances far exceeding the immediate vicinity of the Plateau, consistent with the presence of long-range teleconnections linking the QTP to remote regions of the climate system.
Dashed black lines denote significance thresholds derived from the null model (10th percentile; threshold = 4 for \( F_{\text{out}} \) in \textbf{a} and threshold = 3 for  \( F_{\text{in}} \) in \textbf{b}), used to identify statistically significant nodes.

\clearpage
\begin{figure}[htbp]
  \centering
  \includegraphics[width=1.\textwidth]{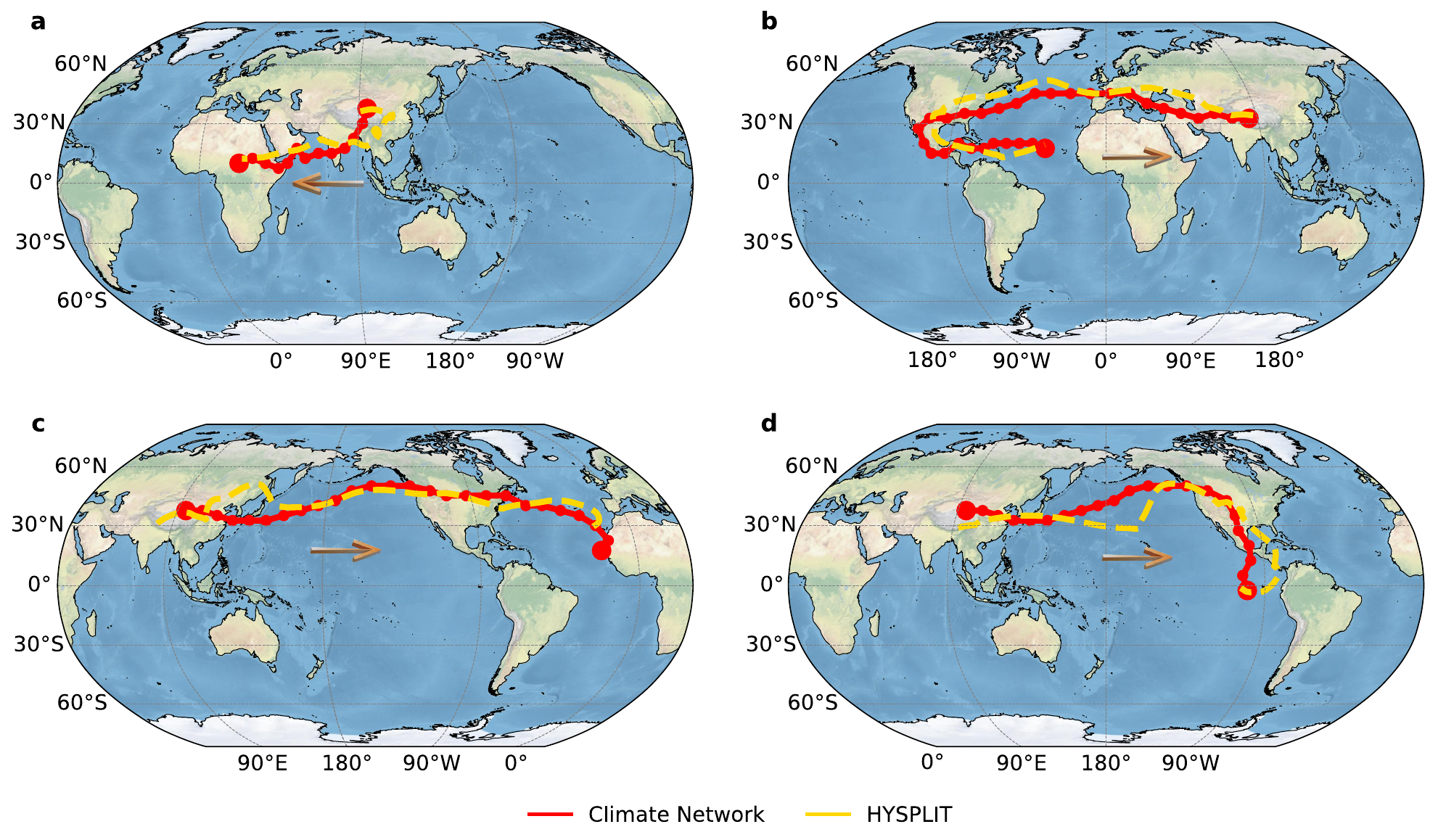}
  \captionsetup{labelformat=empty}
  \caption{\label{SFig_HysplitWithCNforOtherplaces}}
\end{figure} 
\noindent\textbf{Extended Data Fig.~2~\(|\) Teleconnection pathways linking the QTP with the Sahel, Atlantic, and ENSO regions.} \textbf{a}, Outgoing propagation pathway from the QTP to the Sahel, characterized by subsiding airflow associated with the Plateau-scale anticyclonic circulation. \textbf{b}, Incoming pathway from the tropical Atlantic, initially guided by the easterly trade winds and subsequently deflected poleward and eastward through interactions with North Atlantic circulation, including the Gulf Stream region. \textbf{c, d}, Propagation pathways originating from the QTP toward the Atlantic sector and the ENSO regions, primarily mediated by the mid-latitude westerly jet. In all panels, Red curves represent CN-derived optimal propagation paths, while yellow lines indicate HYSPLIT Lagrangian trajectories. 

\clearpage
\begin{figure}[htbp]
  \vspace{-16mm}
  \centering
  \includegraphics[width=1\textwidth]{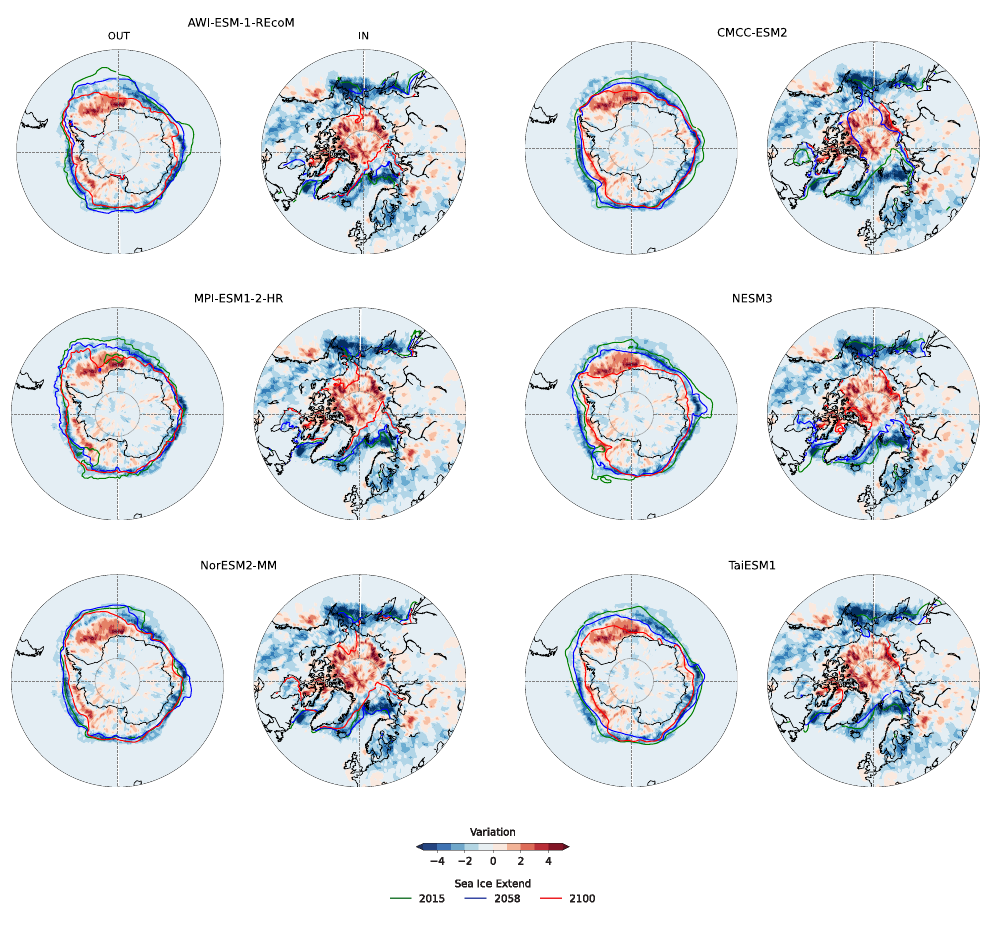}
  \captionsetup{labelformat=empty}
 \caption{\label{SFig_SeaIceBoun}}
\end{figure}
\noindent \textbf{Extended Data Fig.~3~\(|\) Future changes in tripolar teleconnections and their relationship to evolving sea-ice boundaries.} Projected changes in the intensity of QTP–polar teleconnections under future warming, highlighting an increasing concentration of strengthened interactions toward central polar regions. Overlaid contours show boreal-winter sea-ice boundaries, defined by the 15\% sea-ice concentration threshold, derived from multiple CMIP6 models for three representative years: 2015 (green), 2058 (blue), and 2100 (red). The spatial correspondence between regions of enhanced teleconnection strength and retreating sea-ice margins indicates a close association between cryospheric evolution and the reorganization of interhemispheric climate connectivity.

\clearpage
\begin{figure}[htbp]
    \vspace{-5mm}
    \centering
    \begin{minipage}{1\linewidth}
    \centering
    \includegraphics[width=1.\linewidth]{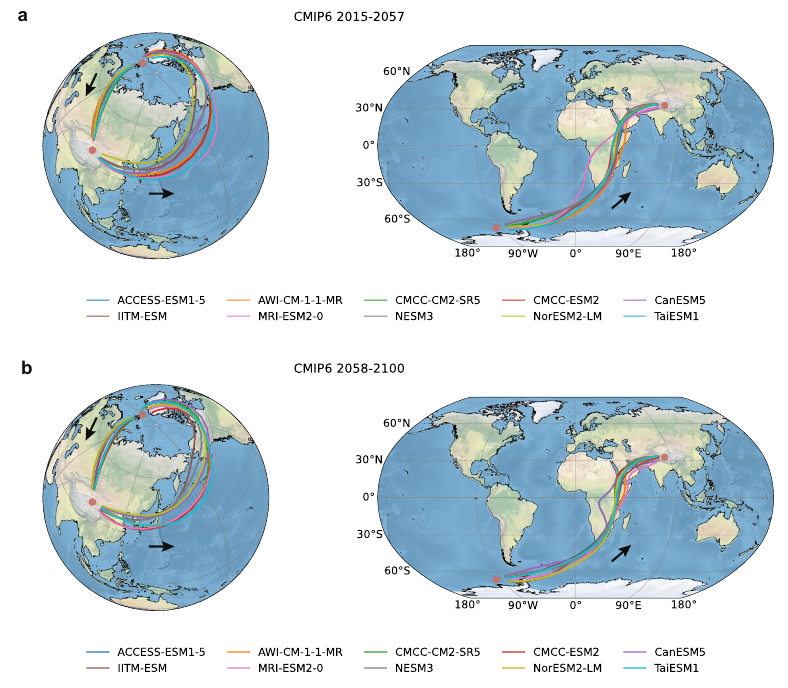}
    \end{minipage}
  \captionsetup{labelformat=empty}
  \caption{\label{SFig_RobustPath_CMIP_Future_2m_ArcAnt}}
\end{figure}
\noindent \textbf{Extended Data Fig.~4~\(|\) Projected robustness of tripolar teleconnection pathways under future warming.} Tripolar teleconnection pathways linking the Qinghai–Tibetan Plateau with the Arctic and Antarctica, derived from climate networks constructed using near-surface air temperature (\textit{tas}) simulated by CMIP6 models under the high-emission SSP5–8.5 scenario. \textbf{a} Tripolar teleconnection structure identified for the early future period (2015–2057). \textbf{b} Corresponding structure for the late future period (2058–2100). The persistence of similar tripolar pathways across datasets indicates that this interaction structure is a robust feature of the climate system rather than a transient or scenario-specific artifact.

\begin{table}[htbp]
    \captionsetup{labelformat=empty}
    \caption{\textbf{Extended Data Table~1~\(|\) Validation of causal pathways within the proposed QTP-Mode.} The table presents the causality tests strictly along the interaction directions defined by the QTP-Mode framework, utilizing both the Liang-Kleeman Information Flow (LKIF) and Granger causality approaches. A checkmark (\checkmark) indicates a statistically significant causal relationship ($P < 0.05$), while a cross ($\times$) denotes no significant causality. A dash (--) indicates that the causality in this specific direction was not evaluated within the core QTP-Mode framework. Note that for Antarctica, both directions are explicitly displayed to highlight its robust unidirectional asymmetry.}
    \label{tab:causality_validation}

    \renewcommand{\arraystretch}{0.9} 
    \setlength{\tabcolsep}{8mm}       
    \begin{tabular}{lccc}
        \toprule
        \toprule 
        \textbf{Region} & \textbf{Interaction Direction} & \textbf{LKIF} & \textbf{Granger Causality} \\
        \midrule
        \addlinespace
        \multicolumn{4}{l}{\textit{\textbf{Tripolar Teleconnection}}} \\
        Arctic     & QTP $\leftrightarrow$ Region & \checkmark & \checkmark \\
        Antarctica & Region $\rightarrow$ QTP     & \checkmark & \checkmark \\
                   & QTP $\rightarrow$ Region     & $\times$   & -- \\
        \multicolumn{4}{l}{\textit{\textbf{Other Tipping Elements}}} \\
        AMOC       & QTP $\leftrightarrow$ Region & \checkmark & \checkmark \\
        ENSO       & QTP $\leftrightarrow$ Region & \checkmark & \checkmark \\
        Sahel      & QTP $\rightarrow$ Region     & \checkmark & \checkmark \\
        Amazon     & Region $\rightarrow$ QTP     & \checkmark & \checkmark \\
        \bottomrule
        \bottomrule
    \end{tabular}
\end{table}

\end{appendices}

\end{document}